# The Influence of Graphene Curvature on Hydrogen Adsorption: Towards Hydrogen Storage Devices

Sarah Goler [a,b], Camilla Coletti [b], Valentina Tozzini [a], Vincenzo Piazza [b], Torge Mashoff [b], Fabio Beltram [a,b], Vittorio Pellegrini [a], and Stefan Heun [a]*

[a] NEST, Institute of Nanoscience – CNR and Scuola Normale Superiore, Piazza San Silvestro 12, 56127 Pisa, Italy

[b] Center for Nanotechnology Innovation @ NEST, Italian Institute of Technology, Piazza San Silvestro 12, 56127 Pisa, Italy

*Corresponding Author: Stefan Heun

Address: NEST, Institute of Nanoscience – CNR and Scuola Normale Superiore, Piazza San Silvestro 12, 56127 Pisa, Italy

Telephone number: +39-050-509472

E-mail address: stefan.heun@nano.cnr.it

The ability of atomic hydrogen to chemisorb on graphene makes the latter a promising material for hydrogen storage. Based on scanning tunneling microscopy techniques, we report on site-selective adsorption of atomic hydrogen on convexly curved regions of monolayer graphene grown on SiC(0001). This system exhibits an intrinsic curvature owing to the interaction with the substrate. We show that at low coverage hydrogen is found on convex areas of the graphene lattice. No hydrogen is detected on concave regions. These findings are in agreement with theoretical models which suggest that both binding energy and adsorption barrier can be tuned by controlling the local curvature of the graphene lattice. This curvature-dependence combined with the known graphene flexibility may be exploited for storage and controlled release of hydrogen at room temperature making it a valuable candidate for the implementation of hydrogen-storage devices.

**Introduction**

Recent years witnessed a rapidly growing commercial interest in renewable energies. There are a number of hurdles to overcome when considering replacing fossil fuels with any renewable form of energy since every step of the process must be optimized. Hydrogen is an attractive possibility since it is abundant in nature and the only byproduct produced during energy consumption is water.[1] However, one pressing issue in developing reliable hydrogen-based technology is solving the issue of hydrogen storage and transport. This paper will focus on this specific obstacle.

Nowadays, the most commonly used storage system –compressed gas– suffers from safety issues linked to the fact that hydrogen is a highly explosive element when under pressure. Furthermore, tanks that can withstand the necessary high pressures are expensive, heavy, and bulky.[2,3] Transporting hydrogen in liquid form, on the other hand, requires both the liquefaction costs and storage at cryogenic temperatures to prevent it from evaporating or building up pressure in a closed container.[4] Clearly, finding a safe mode of transportation and release is essential in pivotal applications such as motor vehicles and in all portable devices.

Several solutions were proposed to address the problem of hydrogen storage using solid-state materials such as metal hydrides, metal organic frameworks, organic chemical hydrides, carbon nanomaterials,[5] and various chemical compounds.[6] However, exploitation of each of these proposals is hindered by some drawbacks. For example, metal hydrides doped with B or Mg offer very high hydrogen gravimetric densities but occupy large volumes.[6] Carbon nanotubes do not meet the gravimetric density standards set by the US Department of Energy (DoE), regarding the use of hydrogen in motor vehicles. Pristine bundled multiwall carbon nanotubes have a hydrogen gravimetric density of 0.52 wt% which can be increased to 2.7 wt% with the creation of nanopores and defects and decoration with Pt nanoparticles.[7] This is still well below the gravimetric density of 5.5wt%, the standard set by the DoE for 2015.[8]

Recently, there has been a renewed interest in carbon-based materials for hydrogen storage thanks to graphene. Graphene was successfully isolated in 2004[9,10] and first considered as a material for hydrogen storage in 2005.[11] The spark came in 2009 when hydrogen-passivated graphene, graphane, was first demonstrated.[12] Graphane is graphene with atomic hydrogen chemically bonded to each graphene lattice atom.[12] Graphane is predicted to be an insulator with a calculated band gap between 3.5eV and 5.4 eV.[13,14] The insulating behavior results from forming C-H bonds, localizing all the delocalized pi-electrons in the graphene lattice. Graphene has taken center stage in the field of hydrogen storage due to its high surface area and vast possibilities of chemical functionalization. A number of theoretical proposals were published including chemically bonding atomic hydrogen to graphene,[13] changing the interplanar distance to form a multilayer graphene nanopump for hydrogen physisorption,[11] and chemically modifying the graphene surface with various transition metals to increase the interaction between molecular hydrogen and graphene.[15-20] The predicted theoretical gravimetric densities are around 5-8 wt%,[6] lower than some of the current materials but well above the standards set by the DoE for 2015.[8] Furthermore, a unique advantage of graphene is the possibility of exploiting its specific flexibility to control the uptake and release of hydrogen at ambient temperature and pressure.[21] Specifically, the idea is based on the increased chemical affinity for hydrogen with convex areas of the graphene surfaces, first theoretically predicted for fullerenes and nanotubes,[22,23] and subsequently on rippled graphene.[24] Consequently, by controlling the curvature and position of ripples one could possibly control the uptake and release of hydrogen.[21]

Experimental validation of these theoretical studies must pass by the study of hydrogenation of rippled graphene structures. Here we present experimental and additional theoretical data ultimately demonstrating the viability of this new concept in which the curvature of graphene is exploited for adsorption and desorption of atomic hydrogen on graphene.

From scanning tunneling microscopy measurements we find that atomic hydrogen attaches to the convexly curved areas of graphene but not to the concave areas indicating that atomic hydrogen does not form stable bonds on concave areas of graphene. Hydrogen attached to the locally convex sites of the lattice is stable up to a temperature of approximately 650°C. When the C-H bonds break, the graphene lattice returns back to that of pristine graphene, *i.e.* the process is reversible. The graphene lattice does not show defects after multiple hydrogenation and dehydrogenation cycles making it a very promising reusable material for hydrogen storage.

**Experimental and Theoretical Methods**

The monolayer graphene samples used in this work were grown by annealing atomically flat 6H-SiC(0001) samples[25] in an induction furnace for 10 minutes at a temperature of about 1450 °C and a pressure of 800 mbar.[26]

Raman emission was excited by means of the 488 nm line of an Ar laser at 1.5-mW power level, focused on the sample to a diffraction-limited sub-micron illumination spot. The details of the setup have been reported elsewhere.[27]

Samples were studied in an ultra high vacuum variable temperature scanning tunneling microscope (STM) with a base pressure of $1 \times 10^{-10}$ mbar. Home etched tungsten tips were degassed in situ, and their oxide was removed by applying a voltage difference of 600V between the tip and a filament and ramping a current through the filament until the electrons emitted from the filament produced an emission current of 10uA measured at the tip. This created stable tips that consistently produced images with atomic resolution. The samples were degassed overnight at around 600 °C to remove adsorbates and water. Images were processed with the WSxM software.[28]

The STM was equipped with a Tectra thermal hydrogen cracker source to produce atomic hydrogen. Monolayer graphene was exposed *in situ* to atomic hydrogen for varying lengths of time. The source was operated with a hydrogen gas flow rate of 1.9 x 10$^{-10}$ liters / second. Using a tabulated cracking efficiency of 100%, this corresponds to an atomic flux of (5.0 ± 0.1) x 10$^{12}$ H atoms/(s cm$^2$).

The DFT calculations were performed using a previously established protocol, shown to be proper for graphene-hydrogenated systems[29] with Kohn_Sham orbitals expanded in plane waves (35 ryd energy cutoff), using Perdew_Burke_Ernzerhof exchange and correlation functionals, Troullier_Martins pseudopotentials and Grimme scheme for Van Der Waals corrections (other details are given in Ref. 21). In Ref. 21, we considered as a model system an orthorhombic cell containing 180 C atoms, laterally compressed to obtain corrugation. Contrary to Ref. 21, in the studies reported here a smaller compression (and corrugation) level was chosen to meet the experimental one. In addition, the STM images were emulated from the integrated electronic charge of the states between the Fermi level and the bias offset, and accounting for the known offset between the Fermi level and the Dirac point. The CPMD3.13 code was used.[30]

**Results**

We study the interaction of hydrogen with monolayer graphene as a function of curvature in samples grown on the silicon face of silicon carbide, SiC(0001). These carbon layers exhibit an innate curvature due to bonds between the buffer layer, a carbon lattice topologically identical to graphene,[27] and the SiC reconstruction whenever the two lattices are in register. Monolayer graphene on SiC(0001) is situated on top of a buffer layer and maintains the superstructure due to the buffer layer's interaction with the SiC reconstruction below. The latter is known as the (6√3x6√3)R30°, but in STM, the superstructure commonly observed is the quasi-(6x6).[31] The corrugation of the monolayer is lower than that of the buffer layer.[32]

The monolayer graphene samples used in this work were first characterized by micro Raman spectroscopy to verify their thickness and homogeneity. After subtraction of the SiC background signal, a typical Raman spectrum averaged on a 12 μm x 12 μm area is reported in Fig. 1a: it shows the characteristic G and 2D bands of monolayer graphene at 1610 cm$^{-1}$ and 2760 cm$^{-1}$. The integrated intensity of the 2D peak as a function of the position on the sample is shown in Fig. 1b and yields the spatial distribution of monolayer (light areas) and buffer layer (dark areas) regions. Data confirm the good homogeneity of the monolayer sample. The shift in the G and 2D bands is a result of doping and strain in the graphene layer.[33,34]

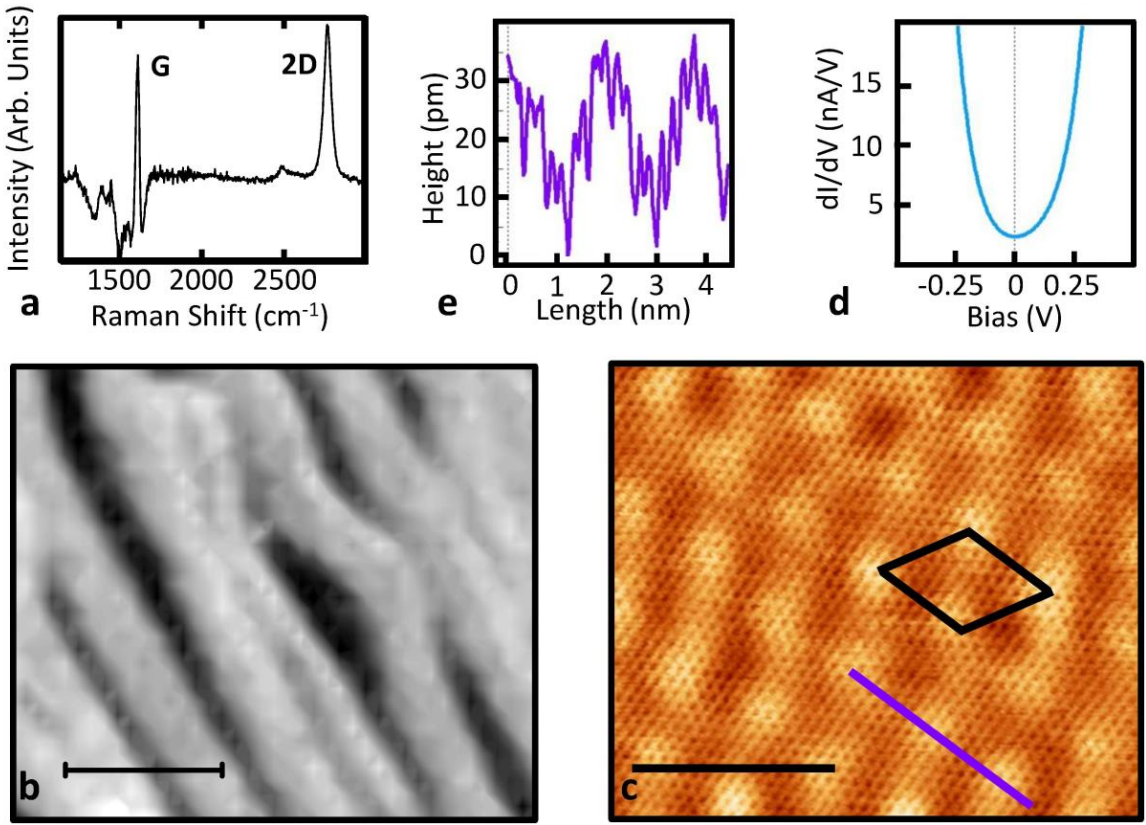

**Figure 1** Characterization of monolayer graphene on SiC(0001). (a) Raman spectrum averaged over a 12 um x 12 um area and after subtraction of the SiC background signal showing the characteristic G and 2D bands at 1610 cm$^{-1}$ and 2760 cm$^{-1}$. (b) Integrated intensity of the 2D peak as a function of the position on the sample. The dark (light) areas show the absence (presence) of the 2D band indicating the absence (presence) of monolayer graphene. Scale bar = 4 um. (c) STM image of monolayer graphene

obtained at bias voltage 115 mV and tunneling current 0.3 nA. The diamond shows the quasi-(6x6) superstructure. Scale bar = 4nm. (d) Average STS data obtained in this area of the sample. (e) Cross section taken along the purple line indicated in (c).

Figure 1c shows an STM image of the clean monolayer graphene surface obtained with a bias voltage of 115mV and a tunneling current 0.3nA. The honeycomb lattice signature of graphene is clearly visible. On a larger scale, the quasi-(6x6) superstructure due to the interaction of the buffer layer with the substrate (black line with diamond shape) is visible too. From the cross section obtained along the purple line in Fig. 1c and plotted in Fig. 1e we measure a peak to peak maximum variation in height of 35pm over a length of about 2nm, along the quasi-(6x6) periodicity. Such a corrugation is typical for monolayer graphene on SiC(0001)[32] and is lower than that of the buffer layer.[27] The STM image shows that the surface is clean with no adsorbates or defects. Along with the STM imaging, scanning tunneling spectroscopy (STS) was performed on the same area. Figure 1d shows an average dI/dV vs. V curve taken from 320 STS curves collected over an evenly-spaced grid on the area where the STM image shown in Fig. 1c was acquired. There were no significant differences observed for spectra on and off the hills. In agreement with other STS studies of monolayer graphene on SiC (0001)[35-37] we observe a minimum at zero bias which does not reach zero and no particular features at the Dirac point (which is located at ca. -0.4 V[38,39]).

Monolayer graphene was exposed *in situ* to atomic hydrogen with an atomic flux of 5 x 10$^{12}$ H atoms/(s cm$^2$) for varying lengths of time and subsequently characterized by STM and STS. At low coverage, *i.e.* after a (5 ± 1) second exposure to atomic hydrogen resulting in a surface coverage of (0.76 ± 0.17)%, the graphene-hydrogen system was stable and it was possible to obtain atomic resolution on the surface. After longer hydrogen exposure times ((25 ± 1) seconds and (145 ± 1) seconds, corresponding to a surface coverage of (3.8 ± 0.2)% and (22 ± 0.6)%, respectively), the system became exceedingly difficult to image and good atomic resolution was not possible. However, STS measurements could be acquired. Figure 2 shows average STS data obtained with atomic hydrogen exposure levels. The black line is from pristine graphene, mirroring the results of previous groups.[35-37] The red line was obtained after hydrogen exposure for 5 seconds and shows a shoulder at negative voltages. A gap of ~0.4eV has formed after a 25 second exposure (green line). After atomic-hydrogen exposure for 145 seconds, the density of states shows that the gap is furthered enlarged. The formation of a gap is evidence of chemisorption of hydrogen on the monolayer. A detailed analysis of the STS data results in a value for the gap of about 1.5eV (see inset to Figure 2). The formation of a gap is expected for graphene with a submonolayer-coverage of hydrogen,[40] while graphane is predicted to have a band gap between 3.5eV and 5.4 eV.[13,14] Again, we did not observe significant differences for spectra on and off the hills. To this end we note that the wavefunction of the chemisorbed hydrogen on graphene is likely to extend more than 1nm from the C-H bond. Since the spacing between maximally convex areas is less than 2nm, it is therefore possible that wavefunctions overlap thus leading to the smoothing of the STS data.

These spectra also explain why STM imaging becomes increasingly difficult for higher hydrogen coverage. Atomic-resolution images of graphene were obtained at voltages below 200mV. With increased hydrogenation, the density of states at those bias voltages decreases to zero and hinders stable tunneling conditions (and therefore images) at those biases.

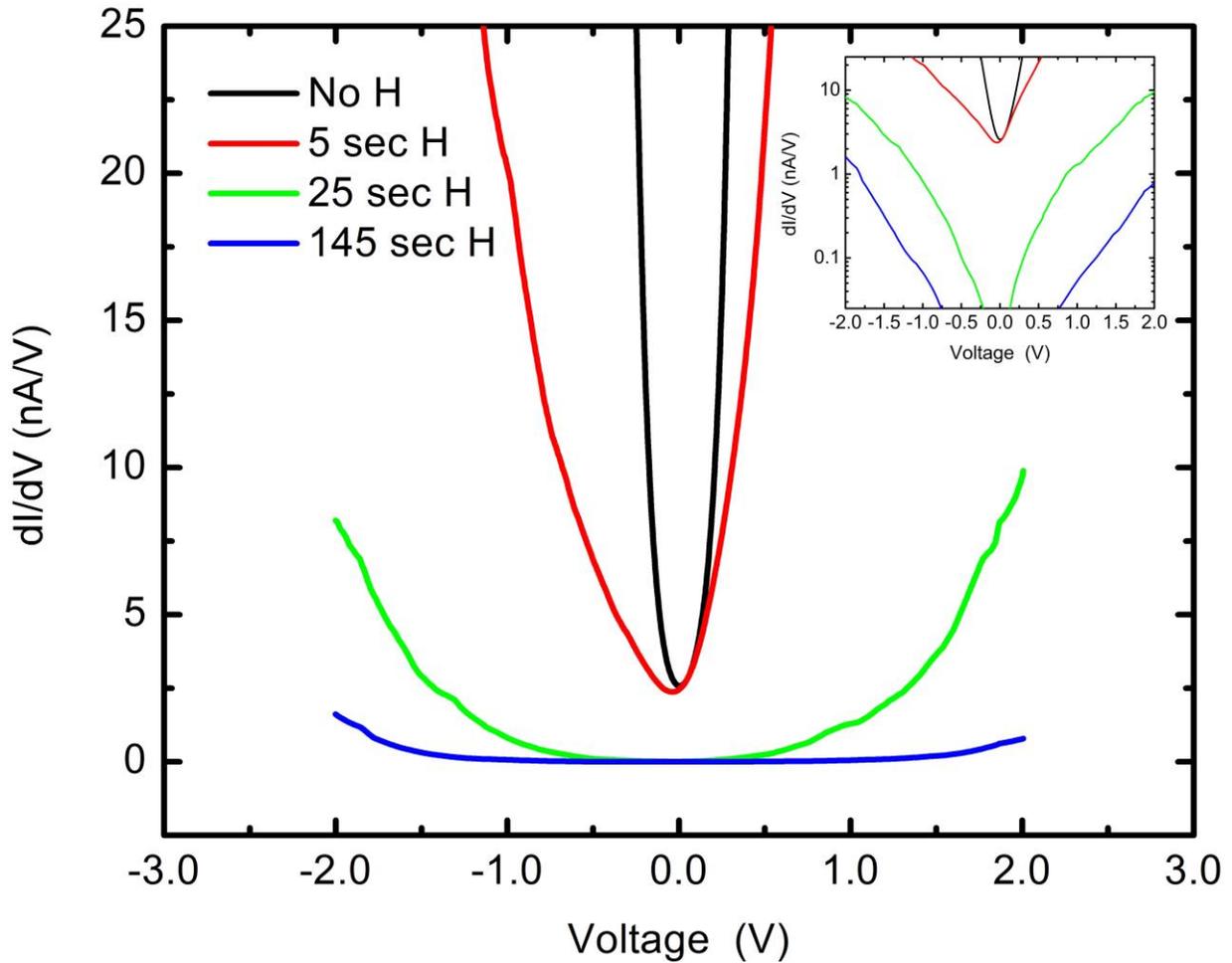

**Figure 2** STS of monolayer graphene upon hydrogen exposure. The black curve corresponds to pristine graphene. The red curve corresponds to a low coverage of hydrogen (5 second hydrogenation). The green and blue curves are for higher coverages (25 seconds and 145 seconds of hydrogenation, respectively). The increasing tendency towards insulating behavior is in agreement with saturating the pi-bonds and opening a gap. The inset shows the same curves plotted on a log scale to clearly show the gap that opens after a 25 second hydrogenation (~0.4eV) and after 145 seconds of exposure to hydrogen (~1.5eV). Setpoints: 0.3nA, 115mV (no H); 0.3nA, 115mV (5 sec H); 0.3nA, 1V (25 sec H); 0.3nA, 2V (145 sec H).

STM images obtained after a 5 second hydrogenation (a representative one is reported in Fig. 4b) show a discernable change from the pristine surface (compare Fig. 4b with Fig. 4a). After hydrogenation, all graphene hills show protrusions at the peaks. Looking closely, it is evident that the protrusions are various combinations of hydrogen atoms. We observe para dimers (Fig. 3a), ortho dimers (Fig. 3c) and tetramers (Fig. 3e). In the para dimer configuration (Fig. 3b inset), the two hydrogen atoms are on opposite sides of the hexagon lattice of graphene. When the two hydrogen atoms bind to neighboring carbon atoms they form an ortho dimer (Fig. 3d inset). The tetramers we observed were formed from two ortho dimers on opposite sides of the hexagonal lattice (Fig. 3f inset). In a 300 nm$^2$ area we counted 3 para- and 3 ortho-dimers as well as 12 tetramers. This corresponds to approximately 1 para- and 1 ortho-dimer in a 10 nm x 10 nm area, while the average number of tetramers found in the same area is about 4.

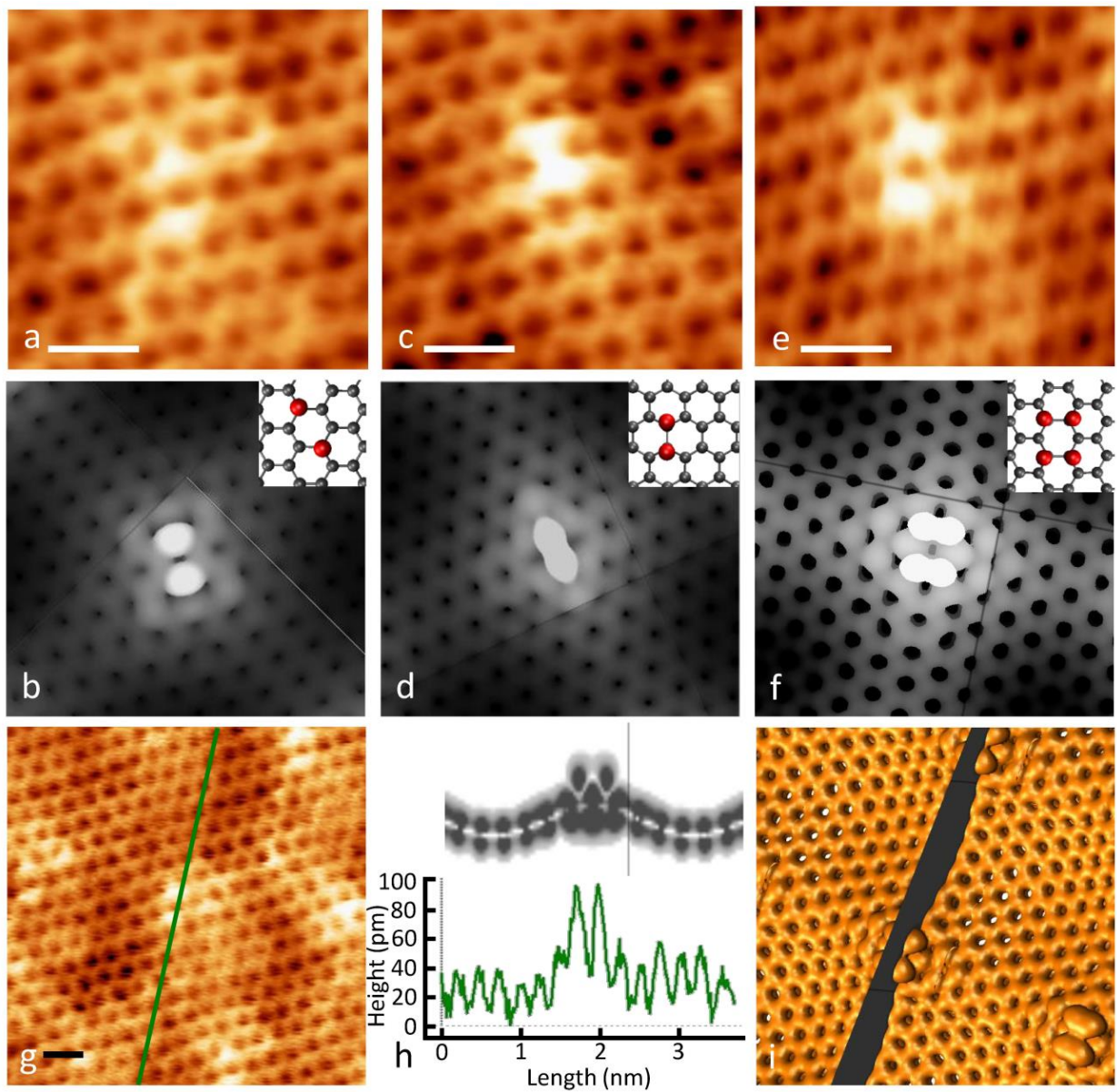

**Figure 3** (a) A para dimer observed in a Scanning Tunneling Microscopy (STM) image obtained on a monolayer graphene surface after a low dose (5 sec) of atomic hydrogen. Scale bar = 4Å. (b) Simulated STM image obtained from DFT calculations of the charge density isosurface of the para dimer. (c) STM image of an ortho dimer (scale bar = 4Å) and (d) corresponding simulation. (e) STM image of a tetramer (scale bar = 4Å.) and (f) corresponding theoretical calculation. Measurement parameters for the STM images were bias voltage 50mV and tunneling current 0.3nA. The insets in b, d, and f are schematics of the various hydrogen configurations observed and described in the respective pairs of panels. (g) Larger scale STM image with a tetramer located in the centre. Scale bar = 4Å. The cross section shown in the lower half of (h) was taken along the green line in (g). The upper half of (h) shows the charge density profile for a tetramer cut along the same orientation as (g) in the charge density map from DFT calculations, represented in (i). In each case (b, d, f and i), the electronic density is evaluated integrating over a sufficient number of electronic states between the Fermi level and the offset. The positions of the atoms and the local curvature where the hydrogen atoms are attached are in agreement. The density charge level for the iso-surfaces and the shades scale are chosen in order to match with experimental images. Higher areas are lighter.

These hydrogen atom conformations were identified also by comparison with Density Functional Theory (DFT) calculations, shown in (Fig. 3b, d, and f). These calculations were performed on a model system, consisting of a supercell of 180 C atoms in which the corrugation is obtained by lateral compression. The calculation setup and model system is the same used in our previous studies,[21] but with a lower level of corrugation, in order to better match the natural curvature of the monolayer. After adding H in an ortho, para, or tetramer conformation on the hills, the system was relaxed and the electronic structure calculated. The simulated STM images were obtained from the iso-electronic density surfaces of the states near the Fermi level (see figure caption). The lighter areas are elevated with respect to the darker areas. Ultimately, the calculated STM images reported in Fig. 3 b, d, and f confirm that what we observe in panels a, c, and e are indeed para dimer, ortho dimer, and tetramer, respectively. Notably, the most prevalent configuration found in our STM scans was the tetramer (Fig. 3e). This might be a result of a high concentration of hydrogen on the surface resulting in complex structural arrangements formed by combining basic dimers, which require the least amount of energy to assemble.  It is also possible that cooperative effects may induce H atoms to cluster on the graphene surface, a process that was theoretically proposed.[21,41]

The bottom panels of Fig. 3 show an STM image (Fig. 3g) and a cross section in green (Fig. 3h) and the theoretical equivalent (Fig. 3i) of a tetramer.  The cross sections of the C-H bonds in the STM images show that hydrogen attaches on top of the hills and forms protrusions of approximately 50pm.  This is much less than the expected C-H bond length, *i.e.* approximately 1.1Å.[13,21] This can be linked to the fact that carbon is slightly more electronegative than hydrogen so that the electronic wavefunction is pulled towards the graphene surface.  The theoretical cross section shows that the hydrogen-carbon bond is concentrated more closely to the carbon atom, indicated by the darker shades of grey that start at approximately half the height of the hydrogen wavefunction (Fig. 3h).  In fact half of 1.1Å is 55pm which agrees quite nicely with the measured change in height of 50pm.

Remarkably, as visible from the STM images in Figs. 3 and 4b, hydrogen preferentially binds on sites where the lattice is maximally convexly curved (*i.e.*, lighter contrast areas).  This is not unexpected since the most favorable formation of an $sp^3$ bonded molecule (such as methane, $CH_4$) is a tetrahedral structure.  In other words, the change from an $sp^2$ to $sp^3$ hybridization requires that the bonds form the lowest energy configuration that deforms the surface towards a tetrahedral form.  If the local curvature is privy to that arrangement, the formation of a C-H bond is more favorable and the barrier for atomic H adsorption is reduced or even eliminated.  This is the same rationale that explains why there is no hydrogen attached in the concavely curved areas of the graphene lattice.

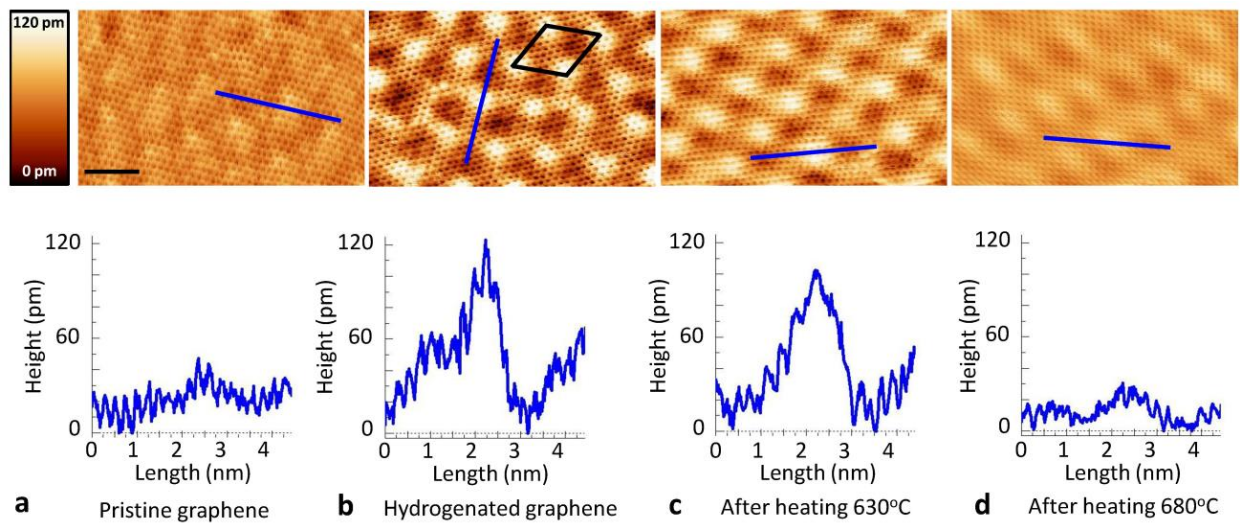

**Figure 4** (a) Scanning Tunneling Microscopy (STM) image of pristine graphene and a cross section below. Bias = 115mV and tunneling current = 0.3nA. (b) STM image of graphene after exposure to atomic hydrogen for 5 seconds resulting in a low coverage of hydrogen. A diamond indicating the quasi-(6x6) superstructure is also shown. The cross section below shows a large increase in corrugation due to the C-H bonds on the convex areas of the graphene surface. Bias = 50mV and tunneling current = 0.3nA. (c) STM image of graphene after annealing for five minutes at 630 °C. The cross section below shows that the hydrogen is still attached to the surface. Bias = 50mV and tunneling current = 0.3nA. (d) STM image of graphene after a five minute annealing at 680 °C showing a clean surface and a corrugation equivalent to that of pristine graphene, indicating that the hydrogen has desorbed from the surface. Bias = 50mV and tunneling current = 0.3nA. The color scale and image size is the same for all STM images. Scale bar = 2nm. All images were obtained at room temperature.

In order to measure the desorption energy barrier for H located on the locally puckered graphene lattice, the hydrogenated-graphene sample was heated in steps of 50°C and subsequently measured by STM. Figure 4 is a summary of the main results obtained. The STM image shown in Fig. 4a was obtained with pristine graphene.  A cross section taken across the surface (blue line) is shown below the STM image and displays a height variation of 40pm, as expected.[32] The average root mean square (RMS) roughness value calculated from this image is 8.97pm.  Following a 5 second exposure to atomic hydrogen, the corrugation dramatically increases reaching a peak-to-peak value of 120pm with an RMS value of 25.6pm (Fig. 4b).  This corrugation remains following a 50°C stepwise 5 minute annealing up to 630°C (Fig 4c).  We note, however, that the bright areas visible on the unheated hydrogenated graphene (Fig. 4b) are not quite restricted to the peaks of the surface reconstruction, but are also visible in the areas with lower local curvature.  After annealing to higher temperatures, on the contrary, hydrogen is found only on the areas of the graphene lattice where the local convex curvature is maximized.  This is in agreement with theory[21] which predicts that the C-H bond energy is greatly diminished when the local curvature becomes concave.  Finally, when the sample is heated to 680°C, hydrogen desorbs from the peaks and the graphene relaxes back to the pristine structure, as shown by the cross section of Fig. 4d.  The RMS value calculated from the image in Fig. 4c (obtained after annealing to 630°C) is 17.86pm. After heating to 680°C (Fig. 4d) it reduces to 7.64pm, similar to the RMS value of the pristine graphene monolayer.  All images in Fig. 4 have the same z-scale to emphasize that hydrogen attaches on the hills, increasing the corrugation along the quasi-(6x6) supercell indicated by the diamond in Fig. 4b which presents the same periodicity found for pristine monolayers.

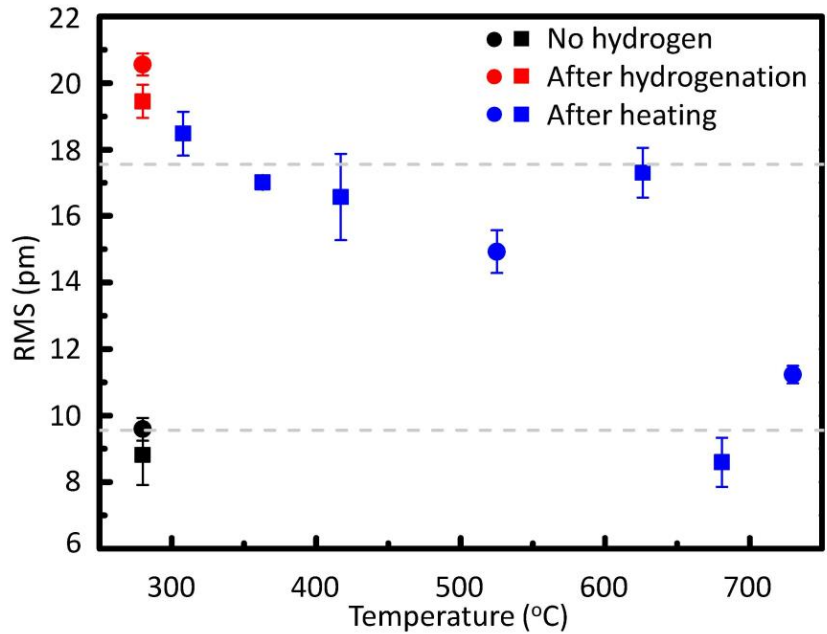

**Figure 5** Average root mean square (RMS) roughness values as a function of temperature. The data was collected from two samples (squares and circles). Each point is the average RMS from the images obtained at that temperature. The error bars denote the standard deviation of these averages. The pristine graphene (black data points) has a low corrugation that dramatically increases when the sample is exposed to a low dose (5 sec) of atomic hydrogen (red data points) at room temperature. The RMS roughness remains high until the sample is heated to 680 °C at which point the corrugation relaxes back to that of pristine graphene. This occurs because the C-H bonds are broken and the hydrogen desorbs from the graphene. The grey dashed lines are guides to the eye.

The corrugation trend is shown in Fig. 5 which provides the RMS values calculated from the height variation of STM images obtained before hydrogenation (black data points), after hydrogenation (red data points) and after progressive heating in steps of 50°C (blue data points). Graphene height variations in the pristine case and after annealing to 680°C are similar. Hydrogenating the sample greatly increases the roughness due to the presence of chemisorbed hydrogen on the surface. This corrugation remains approximately constant up to 630°C. The hydrogen desorbs between 630°C and 680°C. This is confirmed by the measured RMS values that drop back to pristine graphene data in this temperature range.

**Discussion**

A number of studies on hydrogen adsorption and desorption on graphite were reported.[42-47] One such study shows that there are two peaks visible in the thermal desorption spectra at 450K and 560K, 1.4eV and 1.6-1.7eV, respectively.[42] The first peak was attributed to para dimers and the second to ortho dimers. In this case, the surface is natural graphite which is essentially flat whereas in the case of graphene on SiC(0001), the surface is curved. Since convexity stabilizes the adsorbate, desorption from convex areas is expected to occur at higher temperature, in agreement with our results. Moreover, the most stable configuration reported by Dumont *et al.* was the ortho dimer, while in our experiment we saw predominantly tetramers, which can be considered pairs of ortho dimers. In 2009, Balog *et al.* published STM results on monolayer graphene on SiC(0001) showing that hydrogen adsorbs along the superstructure and that hydrogen adsorbs in dimer configurations.[48] The reported extension of the dimers was more than 10Å, and they were not atomically resolved. In a following paper by Ž. Šljivančanin *et al.* from the same group, structures on graphite with the same dimensions were defined as extended hydrogen dimers, configurations with two hydrogen atoms that are not on the same hexagon in the graphene lattice.[46] The study showed both the measured extended dimers covering a length of more than 10Å,[46] which is similar to the work by Balog *et al.*,[48] and their theoretical simulation.[46] In our work, on the contrary, the dimers, two hydrogen atoms on the same hexagon, are observed and atomically resolved directly on top of the carbon atoms. Furthermore, this unequivocal identification was aided by DFT simulations (Fig. 3). We showed that both ortho and para dimers do not extend beyond 4Å (Fig. 3). In addition to the observation of dimers, we observed another stable hydrogen formation which we call tetramers (Fig. 3) not reported in previous studies.

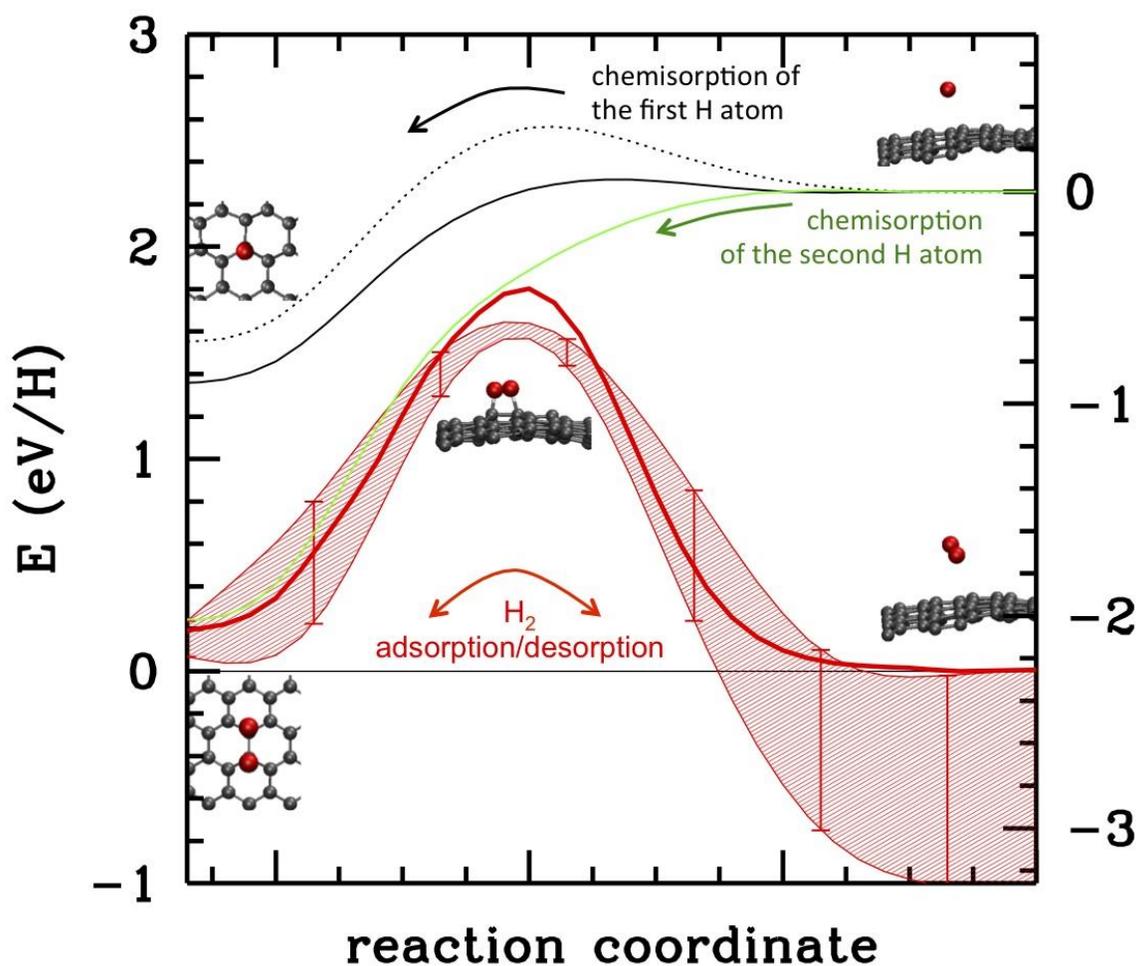

**Figure 6** Energy profiles for the reactions of chemisorption of atomic and molecular hydrogen. Black solid line: chemisorption of a single H atom on a convex site. Green solid line: chemisorption of a second H atom in ortho position with respect to the first. For comparison, curves for the same processes on flat

graphene are reported as dotted lines. Red solid line: associative desorption/dissociative adsorption profile of $H_2$ at 0K temperature. Red shaded band: the same at 300K. The error bars are due to the energy fluctuations of the system. Representative shapshots taken from the simulations are reported. The reaction coordinate is a combination of the H-H and C-H distances (arbitrary units). The reaction path from adsorbed (left) to desorbed hydrogen (right) is followed by constraining this distance to decrease (increase) in a controlled fashion. The energy scales on the left and right y-axis are both in eV, but with two different reference levels: on the left the reference energy level is that of the unbound molecular hydrogen; on the right, the energy level is that of the unbound atomic hydrogen. Their offset is the hydrogen molecule dissociation energy per atom.

We now turn to the discussion of the temperature-dependent data (Figs. 4 and 5). Using 650°C (~930K) as the approximate temperature of the hydrogen desorption from the puckered graphene, the only positions where hydrogen is stable at elevated temperatures, we deduce a desorption energy barrier of 1.4eV. We assume first-order desorption, motivated by the observed "molecule-like" dimer arrangement of the hydrogen atoms on graphene, and similar to what has been observed for hydrogen release from graphite[49] or Rh(110).[50] Then one has $\frac{E_d}{kT_m} = A\tau_m e^{\frac{-E_d}{kT_m}}$,[51] where $\tau_m$ is the time after which the desorption temperature $T_m$ is reached, and $A$ the Arrhenius constant (whose typical value is $10^{13}$ sec$^{-1}$). Using $\tau_m \sim 10^3$ sec (~15 min) one has $E_d/kT_m \sim 33$, hence $E_d$~2.8eV/molecule or 1.4eV/atom. The value obtained for the barrier changes less than 10% with a variation of $A\tau_m$ by one order of magnitude. This value for the desorption barrier is consistent with the DFT calculations (Fig. 6), showing a barrier of 1.55eV at T=0 that decreases to 1.4eV as an effect of the dynamical fluctuations of the graphene sheet at room temperature. Our calculations also show that the dimers are approximately as stable as molecular hydrogen. Two effects contribute to this: first, the chemisorption of an isolated H atom is favored on convex areas (difference between dashed and solid black lines).  The local curvature increases after the first H atom is adsorbed because the carbon atom protrudes out of the graphene plane.[21] This effect induces adhesion of subsequent H atoms (green lines).  The adhesion of atomic hydrogen becomes thus barrierless.  On the other hand, the desorption of hydrogen (or the adsorption of molecules) is governed by energy activation.  Starting from the ortho dimer, the first part of the curve follows the reverse of association, until reaching the activated process, at which point the two H atoms bind together and form a molecule rather than separate, thus following the red line instead of the green one.

These results clearly indicate preferential atomic hydrogen adsorption on locally convex graphene surfaces and the thermal stability of the chemical bond up to 650°C where the local curvature is maximally convex. The atomic hydrogen did not attach to the locally concave parts of the surface, evidence of the instability of the C-H bond at room temperature in these regions.  Our results provide the basis for a graphene based hydrogen storage device that relies on tuning the local curvature to adsorb and release hydrogen at a given temperature and pressure.  Within this approach hydrogen can be adsorbed on the graphene scaffold in regions of convex curvature and stored indefinitely.  To release the hydrogen, the graphene layer can be exposed to an appropriate stimulus that inverts the curvature of the graphene layer releasing the hydrogen.  To achieve the appropriate control of the curvature one possibility could be the use of photoswitching cis/trans molecules attached on one side to a substrate and on the other side to graphene.[21]  By exposing the graphene layer to the appropriate wavelength, the photosensitive molecule will bend or straighten altering the graphene corrugation from concave to convex or vice versa.  Other possible external stimuli for controlling the graphene curvature were described in Ref. 21.

**Conclusions**

We demonstrated the preferential chemisorption of atomic hydrogen on graphene in the areas where the local curvature is maximally convex. This shows that these sites are both the most energetically favorable for hydrogen adsorption and the most stable. We were able to identify various combinations of hydrogen atoms on graphene: para dimers, ortho dimers, and tetramers. We also showed that hydrogen adsorbed on the pi-bonds of the graphene lattice with lower local curvature tends to desorb at a lower temperature, which indicates a lower binding energy in agreement with previous calculations.[21] The curvature-dependent adsorption and desorption of hydrogen provides the basis for the exploitation of graphene as a scaffold for reusable hydrogen storage devices that do not depend on temperature or pressure changes.

**Acknowledgements**

We gratefully acknowledge helpful discussions with Erik Vesselli from the University of Trieste. One of us (V. Pellegrini) acknowledges support from the Italian Ministry of Education, University, and Research (MIUR) through the program ''FIRB – Futuro in Ricerca 2010'' (project title ''PLASMOGRAPH''). CINECA supercomputing center resources were obtained by means of INFM-Progetto di Calcolo Parallelo 2009 and Platform "Computation" of IIT (Italian Institute of Technology).

**References**

1. Crabtree, G. W.; Dresselhaus, M. S. The Hydrogen Fuel Alternative. *MRS Bulletin: Harnessing Materials for Energy* **2008**, *33*, 421–428.

2. Ahluwalia, R. K.; Hua, T. Q.; Peng, J.-K.; Lasher, S.; McKenney, K.; Sinha, J.; Gardiner, M. Technical Assessment of Cryo-Compressed Hydrogen Storage Tank Systems for Automotive Applications. *Int. J. Hydrogen Energy* **2010**, *35*, 4171-4184.

3. U.S. Department of Energy, Energy Efficiency and Renewable Energy, Hydrogen Storage. Last updated 2008.

http://www1.eere.energy.gov/hydrogenandfuelcells/storage/hydrogen_storage.html

4. Ross, D. K. Hydrogen Storage: The Major Technological Barrier to the Development of Hydrogen Fuel Cell Cars. *Vacuum* **2006**, *80*, 1084–1089.

5. Nechaev, Y. S. Carbon Nanomaterials, Relevance to Solving the Hydrogen Storage Problem. *J. Nano Res.* **2010**, *12*, 1-44.

6. Tozzini, V.; Pellegrini, V. Prospects for Hydrogen Storage in Graphene. *Phys. Chem. Chem. Phys.* **2013**, *15*, 80-89.

7. Tsai, P.-J.; Yang, C.-H.; Hsu, W.-C.; Tsai, W.-T.; Chang, J.-K. Enhancing Hydrogen Storage on Carbon Nanotubes via Hybrid Chemical Etching and Pt Decoration Employing Supercritical Carbon Dioxide Fluid. *Int. J. Hydrogen Energy* **2012**, *37*, 6714-6720.

8. U. S. Department of Energy. Energy Efficency and Renewable Energy.

http://www1.eere.energy.gov/hydrogenandfuelcells/storage/current_technology.html

9. Novoselov, K. S.; Geim, A. K.; Morozov, S. V.; Jiang, D.; Zhang, Y.; Dubonos, S. V.; Grigorieva, I. V.; Firsov, A. A. Electric Field Effect in Atomically Thin Carbon Films. *Science* **2004**, *306*, 666-669.

10. Novoselov, K. S.; Jiang, D.; Schedin, F.; Booth, T. J.; Khotkevich, V. V.; Morozov, S. V.; Geim, A. K. Two-Dimensional Atomic Crystals. *Proc. Natl. Acad. Sci. USA* **2005**, *102*, 10451-10453.

11. Patchovskii, S.; Tse, J. S.; Yurchenko, S. N.; Zhechkov, L.; Heine, T.; Seifert, G. Graphene Nanostructures as Tunable Storage Media for Moleculare Hydrogen. *Proc. Natl. Acad. Sci. USA* **2005**, *102*, 10439-10444.

12. Elias, D. C.; Nair, R. R.; Mohiuddin, T. M. G.; Morozov, S. V.; Blake, P.; Halsall, M. P.; Ferrari, A. C.; Boukhvalov, D. W.; Katsnelson, M. I.; Geim, A. K.; et al. Control of Graphene's Properties by Reversible Hydrogenation: Evidence for Graphane. *Science* **2009**, *323*, 610-613.

13. Sofo, J. O.; Chaudhari, A. S.; Barber, G. D. Graphane: A Two-Dimensional Hydrocarbon. *Phys. Rev. B* **2007**, *75*, 153401.

14. Lebègue, S.; Klintenberg, M.; Eriksson, O.; Katsnelson, M. I. Accurate Electronic Band Gap of Pure and Functionalized Graphane from GW Calculations. *Phys. Rev. B* **2009**, *79*, 245117.

15. Lee, H.; Ihm, J.; Cohen, M. L.; Louie, S. G. Calcium-decorated Graphene-Based Nanostructures for Hydrogen Storage. *Nano Lett.* **2010**, *10*, 793-798.

16. Ataca, C.; Aktürk, E.; Ciraci, S.; Ustunel, H. High-Capacity Hydrogen Storage by Metalized Graphene. *Appl. Phys. Lett.* **2008**, *93*, 043123.

17. Beheshti, E.; Nojeh, A.; Servati, P. A First-Principles Study of Calcium-Decorated, Boron-Doped Graphene for High Capacity Hydrogen Storage. *Carbon* **2011**, *49*, 1561-1567.

18. Durgun, E.; Ciraci, S.; Yildirim, T. Functionalization of Carbon-Based Nanostructures with Light Transition-Metal Atoms for Hydrogen Storage. *Phys. Rev. B* **2008**, *77*, 085405.

19. Ao, Z. M.; Peeters, F. M. High-Capacity Hydrogen Storage in Al-Adsorbed Graphene. *Phys. Rev. B* **2010**, *81*, 205406.

20. An, H.; Liu, C.-S.; Zeng, Z.; Fan, C.; Ju, X.  Li-Doped B$_2$C Graphene as Potential Hydrogen Storage Medium.  *Appl. Phys. Lett.* **2011**, *98*, 173101.

21. Tozzini, V.; Pellegrini, V.  Reversible Hydrogen Storage by Controlled Buckling of Graphene Layers. *J. Phys. Chem. C* **2011**, *115*, 25523-25528.

22. Cheng, H.; Cooper, A. C.; Pez, G. P.; Kostov, M. K.; Piotrowski, P.; Stuart, S. J. Molecular Dynamics Simulations on the Effects of Diameter and Chirality on Hydrogen Adsorption in Single Walled Carbon Nanotubes. *J. Phys. Chem. B* **2005**, *109*, 3780-3786.

23. Park, S.; Srivastava, D.; Cho, K. Generalized Chemical Reactivity of Curved Surfaces: Carbon Nanotubes. *Nano Lett.* **2003**, *3*, 1273-1277.

24. Boukhvalov, D. W.; Katsnelson, M. I. Enhancement of Chemical Activity in Corrugated Graphene. *J. Phys. Chem. C* **2009**, *113*, 14176-14178.

25. Frewin, C. L.; Coletti, C.; Riedl, C.; Starke, U.; Saddow, S. E. A Comprehensive Study of Hydrogen Etching on the Major SiC Polytypes and Crystal Orientations. *Mater. Sci. Forum* **2009**, *615-617*, 589-592.

26. Forti, S.; Emtsev, K. V.; Coletti, C.; Zakharov, A. A.; Starke, U. Large-Area Homogeneous Quasifree Standing Epitaxial Graphene on SiC(0001): Electronic and Structural Characterization. *Phys. Rev. B* **2011**, *84*, 125449.

27. Goler, S.; Coletti, C.; Piazza, V.; Pingue, P.; Colangelo, F.; Pellegrini, V.; Emtsev, K. V.; Forti, S.; Starke, U.; Beltram, F.; et al.  Revealing the Atomic Structure of the Buffer Layer Between SiC(0001) and Epitaxial Graphene.  *Carbon* **2013**, *51*, 249-254.

28. Horcas, I.; Fernandez, R.; Gomez-Rodriguez J. M.; Colchero, J.; Gomez-Herrero, J.; Baro A. M. WSXM: a Software for Scanning Probe Microscopy and a Tool for Nanotechnology. *Rev. Sci. Instrum.* **2007**, *78*, 013705.

29. Tozzini, V.; Pellegrini, V. Electronic Structure and Peierls Instability in Graphene Nanoribbons Sculpted in Graphane. *Phys. Rev. B* **2010**, *81*, 113404.

30. CPMD, http://www.cpmd.org/, Copyright IBM Corp 1990- 2008, Copyright MPI für Festkörperforschung Stuttgart 1997-2001.

31. Riedl, C.; Coletti, C.; Starke, U. Structural and Electronic Properties of Epitaxial Graphene on SiC(0001): a Review of Growth, Characterization, Transfer Doping and Hydrogen Intercalation. *J. Phys. D: Appl. Phys.* **2010**, *43*, 374009.

32. Varchon, F.; Mallet, P.; Veuillen, J.-Y.; Magaud, L. Ripples in Epitaxial Graphene on the Si-Terminated SiC(0001). *Phys. Rev. B* **2008**, *77*, 235412.

33. Das, A.; Pisana, S.; Chakraborty, B.; Piscanec, S.; Saha, S. K.; Waghmare, U. V.; Novoselove, K. S.; Krishnamurthy, H. R.; Geim, A. K.; Ferrari, A. C.; et al. Monitoring Dopants by Raman Scattering in an Electrochemically Top-Gated Graphene Transistor. *Nat. Nanotechnol.* **2008**, *3*, 210-215.

34. Röhrl, J.; Hundhausen, M.; Emtsev, K. V.; Seyller, Th.; Graupner, R.; Ley, L. Raman Spectra of Epitaxial Graphene on SiC(0001). *Appl. Phys. Lett.* **2008**, *92*, 201918.

35. Brar, V. W.; Zhang, Y.; Yayon, Y.; Ohta, T.; McChesney, J. L.; Bostwick, A.; Rotenberg, E.; Horn, K.; Crommie, M. F. Scanning Tunneling Spectroscopy of Inhomogeneous Electronic Structure in Monolayer and Bilayer Graphene on SiC. *Appl. Phys. Lett.* **2007**, *91*, 122102.

36. Lauffer, P.; Emtsev, K. V.; Graupner, R.; Seyller, Th.; Ley, L.; Reshanov, S. A.; Weber, H. B. Atomic and Electronic Structure of Few-Layer Graphene on SiC(0001) Studied with Scanning Tunneling Microscopy and Spectroscopy. *Phys. Rev. B* **2008**, *77*, 155426.

37. Rutter, G. M.; Guisinger, N. P.; Crain, J. N.; Jarvis, E. A. A.; Stiles, M. D.; Li, T.; First, P. N.; Stroscio, J. A. Imaging the Interface of Epitaxial Graphene with Silicon Carbide via Scanning Tunneling Microscopy. *Phys Rev. B* **2007**, *76*, 235416.

38. Ohta, T.; Bostwick, A.; McChesney, J. L.; Seyller, T.; Horn, K.; Rotenberg, E. Interlayer Interaction and Electronic Screening in Multilayer Graphene Investigated with Angle-Resolved Photoemission Spectroscopy. *Phys. Rev. Lett.* **2007**, *98*, 206802.

39. Coletti, C.; Riedl, C.; Lee, D. S.; Krauss, B.; Patthey, L.; von Klitzing, K.; Smet, J. H.; Starke, U. Charge Neutrality and Band-Gap Tuning of Epitaxial Graphene on SiC by Molecular Doping. *Phys. Rev. B* **2010**, *81*, 235401.

40. Duplock, E. J.; Scheffler, M.; Lindan, P. J. D. Hallmark of Perfect Graphene. *Phys. Rev. Lett.* **2004**, *92*, 225502.

41. Stojkovic, D.; Zhang, P.; Lammert, P. E.; Crespi, V. H. Collective Stabilization of Hydrogen Chemisorptions on Graphenic Surfaces. *Phys. Rev. B.* **2003**, *68*, 195406.

42. Dumont, F.; Picaud, F.; Ramseyer, C.; Girardet, D.; Ferro, Y.; Allouche, A. Model for Thermal Desorption of Hydrogen Atoms from a Graphite Surface Based on Kinetic Monte Carlo Simulations. *Phys. Rev. B* **2008**, *77*, 233401.

43. Hornekær, L.; Šljivančanin, Ž.; Xu, W.; Otero, R.; Rauls, E.; Stensgaard, I.; Lægsgaard, E.; Hammer, B.; Besenbacher, F. Metastable Structures and Recombination Pathways for Atomic Hydrogen on the Graphite(0001) Surface. *Phys. Rev. Lett.* **2006**, *96*, 156104.

44. Hornekær, L.; Xu, W.; Lægsgaard, E.; Besenbacher, F. Long Range Orientation of Meta-Stable Atomic Hydrogen Adsorbate Clusters on the Graphite(0001) Surface. *Chem. Phys. Lett.* **2007**, *446*, 237-242.

45. Zecho, T.; Güttler, A.; Sha, X.; Jackson, B.; Küppers, J. Adsorption of Hydrogen and Deuterium Atoms on the (0001) Graphite Surface. *J. Chem. Phys.* **2002**, *117*, 8486-8492.

46. Šljivančanin, Ž.; Rauls, E.; Hornekær, L.; Xu, W.; Besenbacher, F.; Hammer, B. Extended Atomic Hydrogen Dimer Configurations on the Graphite(0001) Surface. *J. Chem. Phys.* **2009**, *131*, 084706.

47. Hornekær, L.; Rauls, E.; Xu, W.; Šljivančanin, Ž.; Otero, R.; Stensgaard, I.; Lægsgaard, E.; Hammer, B.; Besenbacher, F. Clustering of Chemisorbed H(D) Atoms on the Graphite(0001) Surface due to Preferential Sticking. *Phys. Rev. Lett.* **2006**, *97*, 186102.

48. Balog, R.; Jørgensen, B., Wells, J.; Lægsgaard, E.; Hofmann, P.; Besenbacher, F.; Hornekær, L. Atomic Hydrogen Adsorbate Structures on Graphene. *J. Am. Chem. Soc.* **2009**, *131*, 8744-8745.

49. Denisov, E. A.; Kompaniets, T. N. Kinetics of Hydrogen Release from Graphite after Hydrogen Atom Sorption. *Phys. Scr.* **2001**, *T94*, 128 – 131.

50. Vesselli, E.; Campaniello, M.; Baraldi, A.; Bianchettin, L.; Africh, C.; Esch, F.; Lizzit, S.; Comelli, G. A Surface Core Level Shift Study of Hydrogen-Induced Ordered Structures on Rh(110). *J. Phys. Chem. C* **2008**, *112*, 14475 – 14480.

51. Woodruff, D. P.; Delchar, T. A. Modern Techniques of Surface Science. *Cambridge University Press* **1994**.

**Table of Contents Image**

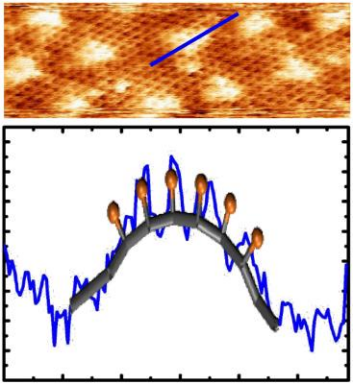